# Implementing Effective Changes in Software Projects to Optimize Runtimes and Minimize Defects

Kartik Gupta
kartikgupta@outlook.com

*Abstract*—The continuous evolution of software projects necessitates the implementation of changes to enhance performance and reduce defects. This research explores effective strategies for learning and implementing useful changes in software projects, focusing on optimizing runtimes and minimizing software defects. A comprehensive review of existing literature sets the foundation for understanding the current landscape of software optimization and defect reduction. The study employs a mixedmethods approach, incorporating both qualitative and quantitative data from software projects before and after changes were made. Key methodologies include detailed data collection on runtimes and defect rates, root cause analysis of common issues, and the application of best practices from successful case studies. The research highlights critical techniques for learning from past projects, identifying actionable changes, and ensuring their effective implementation. In-depth case study analysis provides insights into the practical challenges and success factors associated with these changes. Statistical analysis of the results demonstrates significant improvements in runtimes and defect rates, underscoring the value of a structured approach to software project optimization. The findings offer actionable recommendations for software development teams aiming to enhance project performance and reliability. This study contributes to the broader understanding of software engineering practices, providing a framework for continuous improvement in software projects. Future research directions are suggested to refine these strategies further and explore their application in diverse software development environments.

## I. Introduction

Business users are demanding tools that support businesslevel interpretations of their data. At a panel on software analytics at ICSE'12, industrial practitioners lamented the state of the art in software analytics [1]. Panelists commented "prediction is all well and good, but what about decision making?". Note that these panelists were more interested in the interpretations and follow-up that occurs after the mining, rather than just the mining itself. So:

- Instead of just accepting *predictions* on how many software defects to expect, business users might now demand a *plan* to reduce the likelihood of those defects.
- Instead of just accepting *predictions* on the runtime time of their software, business users might now demand a *plan* to reduce that runtime.

In response to this business-level demands for planners, I propose a novel *planning* method called XTREE for learning changes to a software system such that its performance "improves", according to some measure. This paper uses XTREE to reduce the expected value of the defects in Jureczko et al.'s JAVA systems [2]; and the runtimes in software configured by Siegmund et al. [3].

The contributions of this paper are (1) the new XTREE algorithm and (2) an evaluation strategy for planners. In the case studies of this paper my evaluations shows XTREE performing significantly "better" than planners proposed in my prior work [4], [5], where "better" means:

- *Effective:* i.e. if a plan is applied then some statistically significant change in expected values of the results will be observed.
- *Succinct:* i.e. given two plans with the same effectiveness, I prefer the shorter one. Such shorter plans are easier to explain (less to show) and faster to implement (less to change).
- *Surprising*; i.e. I want the planners to (occasionally) tell us things I do not know already (otherwise, there would be no point to using the planner).

The rest of this paper is describes my data, my planners, and the experiment that ranks XTREE against alternate approaches. This is followed by notes on related work and validity. Note that, to allow for reproducibility, all scripts and data used in this study are available on-line at github.com/aise/XTREE.

## II. Preliminaries A.

*What is a "Plan"?*

my planners use tables of data with independent features and one dependent feature, called the "class". Classes are weighted such that they indicate what rows are "good" and what rows are "bad". Plans change a row such that it is more like they will be more similar to "good" than "bad". Specifically, for every test example $Z$, planners proposes a plan $\Delta$ to adjust feature $Z_j$:

$$\forall \delta_j \in \Delta : Z_j = Z_j + \delta_j$$

For example, to simplify a large bug-prone method, my planners might suggest to a developer to reduce its size (i.e. refactor that code by, say, splitting it across two simpler functions).

Note that I make no assumption that a plan mentions every features (so plan1 can be more succinct than plan2 when plan1 mentions fewer features than plan2).

### B. From Prediction to Planning

This paper is about the next step *after* prediction. Suppose a business user is presented a prediction and they do not like what they see; e.g. the runtimes are too long of the number of defects is too high. This user may then ask a *planning* question; i.e. "what can I change to do better than that?".

Before exploring automatic methods to answer the planning question, I first comment on two manual methods.

One way to propose changes to a project would be to ask some smart experienced person for their opinion on how to (e.g.) reduce defects and/or decrease runtimes. Sometimes such advice is an effective strategy and sometimes it is not. According to Passos et al. [6], developers may often generalize lessons learned from a few past projects to all future endeavors without carefully considering the context. As they point out, "past experiences were taken into account without much consideration for their context" [6]. Jørgensen & Gruschke [7] similarly caution that many software engineering "gurus" fail to use past experiences to improve their future decision-making, and misguided advice from previous projects can be detrimental to new ones [7]. Thus, I advocate for a "trust, but verify" approach. When a software expert offers guidance, it is prudent to consult additional sources to see if better alternatives exist, as a sanity check.

Another approach to identify project changes is to rely on the peer review mechanisms employed by the software engineering (SE) research community. This method proposes alterations that align with internationally accepted best practices. However, two issues arise with this approach. First, due to the rapid advancements in software engineering, there may be instances where there are no widely accepted best practices. For example, predicting software runtimes based on Makefile choices has only recently been explored. While the work by Siegmund et al. [3] represents state-of-the-art research, it is too new to be considered an internationally accepted best practice.

Second, given the diversity of SE products, practices, and teams, the project in question may differ significantly from prior work. Recent research in *local learning* compares (1) global models learned from all available data with (2) localized models derived from clusters within the data [8]–[14]. A recurring finding is that localized models often result in better predictions for both effort and defects, yielding more accurate median outcomes and reduced prediction variance.§V-B3 of this paper offers yet another locality result:

- One standard rule in the literature is that it is useful to implement modules such that they are internally cohesive (use much of their own local methods) while being loosely coupled with other classes [15].
- While that may be true in general, for particular classes other changes may be more important (later in this paper, I show one set of results were that is indeed the case).

In summary, it is useful to have automatic methods to recommend changes. Such methods can fill in for human gurus (if such gurus are absent) or to offer a second opinion. Also, prior to making automatic recommendations, it is wise to first stratify the data (clump it into related examples) then generate advice specific to each clump. Accordingly, the rest of this paper defines and evaluates automatic methods to find plans from $N$ examples divided into many clumps.

*C. Trusting the Changes*

XTREE is evaluated by comparing predicted performance scores before and after a planner makes changes to the feature values of an example: After making those changes, I may have a new example that has never seen before. Therefore, it must be asked *"can I trust the predictions made on such new examples?"*

To answer this question, I note that data miners explore two "clouds" of data: (1) the cloud of training examples and (2) the cloud of test examples. For a visualization of these clouds, see Figure 2.

I should mistrust the predictions made by a model if it is being applied to examples that are too far away from the training cloud. To test for "too far", I can run a data mining experiment that tests how well a model learned from the training data applies to the test data. Such experiments return some performance value.

Note that predictions about changes that fall within the space of the training+test data, will be at least as reliable as the performance value found in the above data mining experiment. Given this, one thing can be asserted about predictions on changed examples:

- Predictions for changes that move examples towards/away from the training data can be trusted more/less (respectively).

Accordingly, I should use *trust-increasing* planners that generate changed examples *closer* to the training examples. To see how this works, Figure 2 is from the *ivy* data set, which is one of the Jureczko data sets explored in this paper. It shows: (1) the training examples in gray, (2) the test examples in red, and (3) the changed examples displaced after applying a plan (in green). Note that the the changed examples cases (shown in green) fall closer to the training cases (shown in gray) than the test cases (shown in red).

In that green region of changed examples, my belief in the value of predictions will be as much (or more) as my belief in the value of the predictions in the red region (that contains the original test data). This pattern of Figure 2 (where the changes examples are found closer to the training cases than the test cases) has been observed in all the other data sets studied in this paper. Hence, I can assert that predictors learned from these training examples have some authority in the regions contain the changes examples.

That said, the above comes with some important caveats:

- I strongly recommend that predictors are assessed prior to planning. That issue is explored further in the §III.
- Planners should be designed to be *trust increasing*. I list four such planning methods in §IV.
- Where possible, planners should be assessed via some external oracle that can accurately assess new examples. For an example of that kind of analysis, see §VI.

| | | |
|---|---|---|
| amc | average method complexity | e.g. number of JAVA byte codes |
| avg cc | average McCabe | average McCabe's cyclomatic complexity seen in class |
| ca | afferent couplings | how many other classes use the specific class. |
| class. | | |
| cam | cohesion amongst classes | summation of number of different types of method parameters in every method divided by a multiplication of number of different method parameter types in whole class and number of methods. |
| cbm | coupling between methods | total number of new/redefined methods to which all the inherited methods are coupled |
| cbo | coupling between objects | increased when the methods of one class access services of another. |
| ce | efferent couplings | how many other classes is used by the specific class. |
| dam | data access | ratio of the number of private (protected) attributes to the total number of attributes |
| dit | depth of inheritance tree | |
| ic | inheritance coupling | number of parent classes to which a given class is coupled (includes counts of methods and variables inherited) |
| lcom | lack of cohesion in methods | number of pairs of methods that do not share a reference to an case variable. |
| locm3 | another lack of cohesion measure | if $m,a$ are the number of $methods, attributes$ in a class number and $\mu(a)$ is the number of methods accessing an attribute, then $lcom3 = ((\frac{1}{a}\sum_{j}^{a}\mu(a,j)) - m)/(1 - m)$. |
| loc | lines of code | |
| max cc | maximum McCabe | maximum McCabe's cyclomatic complexity seen in class |
| mfa | functional abstraction | number of methods inherited by a class plus number of methods accessible by member methods of the class |
| moa | aggregation | count of the number of data declarations (class fields) whose types are user defined classes |
| noc | number of children | |
| npm | number of public methods | |
| rfc | response for a class | number of methods invoked in response to a message to the object. |
| wmc | weighted methods per class | |
| nDefects | raw defect counts | Numeric: number of defects found in post-release bug-tracking systems. |
| defect | defects present? | Boolean: if $nDefects > 0$ then $true$ else $false$ |

Fig. 1. OO measures used in our defect data sets. Last lines, shown in denote the dependent variables.

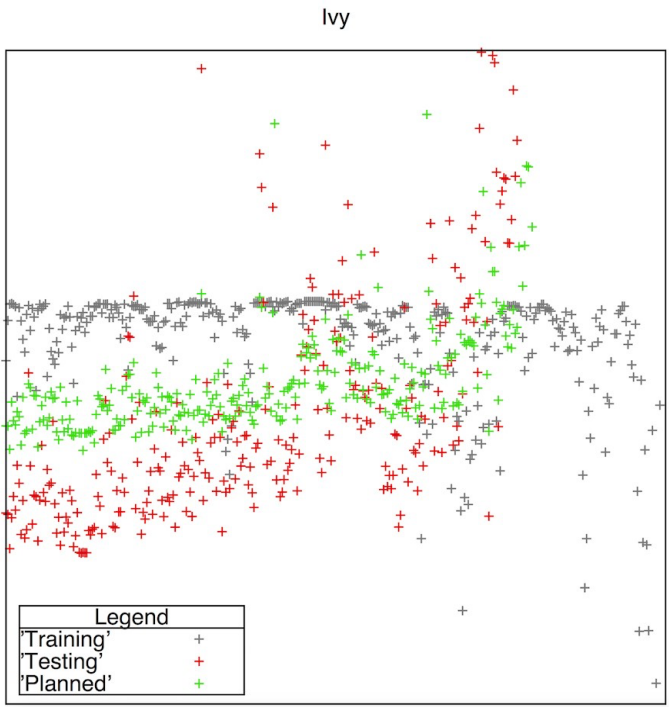

| data set | cases | % defective |
|---|---|---|
| ant | 947 | 22 |
| camel | 1819 | 19 |
| jedit | 1257 | 2 |
| ivy | 352 | 11 |
| log4j | 244 | 92 |
| lucebe | 442 | 59 |
| poi | 936 | 64 |
| synapse | 379 | 34 |
| velocity | 410 | 34 |
| xalan | 2411 | 99 |
| xerces | 1055 | 74 |

Fig. 3.

| Project | Lang. | Features |
|---|---|---|
| BDBC: Berkeley DB | | |
| BDBJ: Berkeley DB | | |
| Apache | | |
| SQLite | | |
| LLVM | | |
| x264 | | |

Jureczko data: columns in the format of Figure 1.

| Domain | | LOC | | Config |
|---|---|---|---|---|
| Database | C | 219,811 | 18 | 2560 |
| Database | Java | 42,596 | 32 | 400 |
| Web Server | C | 230,277 | 9 | 192 |
| Database | C | 312,625 | 39 | 3,932,160 |
| Compiler | C++ | 47,549 | 11 | 1024 |
| Video Enc. | C | 45,743 | 16 | 1152 |

Fig. 4. Siegmund data. For SQLite, the data contains 4,553 configurations for prediction modeling and 100 additional random configurations for prediction evaluation, see [16].

class, where the classes are described in terms of nearly two dozen metrics such as number of children, lines of code, etc.

Fig. 2.

## III. TEST DATA

To assess my planning methods, I use data from Jureczko et al.'s object-oriented JAVA systems [2] and software system configuration data from by Siegmund et al. [3]. See github.com/ai-se/XTREE#data for full access to this data.

The Jurecko data records number of known defects for each For details on the Jurecko data, see Figure 1 and Figure 3. For the most part, the methods of this paper treat the Jurecko as a discrete class data set, where *defects* are true if the raw defect count is greater than zero. The one exception will be the 'best-in-cluster" method that reflects on the numeric value of the raw defect count (see §IV-A3).

The Siegmund data, described in Figure 4, records the runtimes of compiled systems. To make that data, Siegmund et al. perturbed the configuration parameters in the Makefiles of six systems: Apache, SQLite, LLVM, x264 and two versions of the Berkeley database (one written in "C" and one in Java). Figure ?? shows an example of a feature model defining valid combinations of settings to on the the Siegmund et al. datasets.

As mentioned in the last section, this approach depends on having effective predictors for assessing the results. For the Siegmind data, this criteria was relatively easy to achieve. The data in those data sets have a continuous class (runtime of the compiled system) so the performance of a quality predictor can be measured in terms of difference between the predicted runtime $p$ of test case items and their actual runtimes $a$ using $s = 1 - \frac{abs(a-p)}{a}$ (and *higher* values are *better*). This paper explores six Siegmund configuration data sets: Berkeley DB (Java and C versions), Apache, SQLite, LLVM, and x264. As a preliminary study, I split that data into equal sized train:test groups and trained a Random Forest Regressor (from the SciKit learn kit [17]) on one half, then applied to the other. This achieved nearly perfect scores of $s = \{99.9, 99.8, 99.4, 99.1, 96.1\}$%. That is, I can be very confident that the predictors from the Siegmund data can assess my plans. (Aside: if the reader doubts that such high scores are achievable, I note that these scores are consistent with those achieved by predictors built by Siegmund et al. [3].) It proved to be more complicated to commission the Jureczko data sets for this study. For that data, I found that the quality predictors built from this data are far from perfect; However, for some

| data set | Data set properties | | | | | Results from learning | | | | | | | | |
|---|---|---|---|---|---|---|---|---|---|---|---|---|---|---|
| | training | | testing | | | untuned | | | tuned | | | change | | |
| | versions | cases | versions | cases | % defective | pd | pf | good? | pd | pf | good? | pd | pf | |
| jedit ivy | 3.2, 4.0, 4.1, 4.2 | 1257 | 4.3 | 492 | 2 | 55 | 29 | | 64 | 29 | y | 9 | 0 | |
| camel | 1.1, 1.4 | 352 | 2.0 | 352 | 11 | 65 | 35 | y | 65 | 28 | y | 0 | -7 | |
| ant | 1.0, 1.2, 1.4 | 1819 | 1.6 | 965 | 19 | 49 | 31 | | 56 | 37 | | 5 | 6 | |
| synapse | 1.3, 1.4, 1.5, 1.6 | 947 | 1.7 | 745 | 22 | | | | | | | | | |
| velocity | 1.0, 1.1 | 379 | 1.2 | 256 | 34 | 49 | 13 | y | 63 | 16 | y | 14 | 3 | |
| lucene | 1.4, 1.5 | 410 | 1.6 | 229 | 34 | 45 | 19 | | 47 | 15 | | 2 | -4 | |
| poi | 2.0, 2.2 | 442 | 2.4 | 340 | 59 | 78 | 60 | | 76 | 60 | | -2 | 0 | |
| xerces | 1.5, 2, 2.5 | 936 | 3.0 | 442 | 64 | 56 | 25 | | 60 | 25 | y | 4 | 0 | |
| log4j | 1.0, 1.2, 1.3 | 1055 | 1.4 | 588 | 74 | 56 | 31 | | 60 | 10 | y | 4 | -21 | |
| xalan | 1.0, 1.1 | 244 | 1.2 | 205 | 92 | | | | | | | | | |
| | 2.4, 2.5, 2.6 | 2411 | 2.7 | 909 | 99 | 30 | 31 | | 40 | 29 | | 10 | -2 | × |
| | | | | | | 32 | 6 | | 30 | 6 | | -2 | 0 | × |
| | | | | | | 38 | 9 | | 47 | 9 | | 9 | 0 | × |

Fig. 5. Training and test *data set properties* for Jureczko data, sorted by % defective examples. On the right-hand-side, we show the *results from learning*. Data is "good" if it has recall over 60% and false alarm under 40% (and note that, after tuning, there are more "good" marked with " " before). Results show large improvements in performance, after tuning (lower *pf* or higher *pd*). The data in the three bottom rows, marked with "×", are "not good" since their test suites have so few non-defective examples (less than 5% of the total sample) that it becomes harder to find better data towards which we can displace test data.

These feature models were used by Siegmund et al. to ensure all their perturbations are value (I will use the same models to cull invalid plans). Given those valid perturbations, the systems were then compiled and Siegmund et al. recorded how long each perturbation took to run a test suite.

My evaluation strategy (discussed below) divides this data into a training a test set. From the train set I apply a data miner (to learn a quality predictor) and various planning methods (to learn different plans). Next, I try applying each of those plans to the test set and ask the quality predictor to assess the changed examples. Finally, I say that the "best" planner is the one that most reduces the predicted values in the changed examples.

data sets, the predictors could be salvaged using the techniques discussed in this section.

Figure 5 shows my preliminary studies with the Jureczko data. Given access to $V$ released versions, I test on version $V$ and train on the available data from $V-1$ earlier releases (as shown in Figure 5, this means that I are training on hundreds to thousands of classes and testing on smaller test suites). Note the three bottom rows marked with ×: these contain predominately defective classes (two-thirds, or more). In such data sets, it is hard to distinguish good from bad (since there are so many bad examples).

The Jureczko data uses non-numeric discrete classes ("defective" or "not"). For such data, quality predictor is be measured using (1) the probability of detection (a.k.a. "pd" or recall): the percent of faulty classes in the test data detected by the *predictor*; and (2) the probability of false alarm (a.k.a. "pf"): the percent of non-fault classes that are *predicted* to be defective.

As a preliminary study, I split the Jureczko data into equal sized train:test groups. Random Forests (again, from the SciKit learn kit [17]) were built from the training data, then applied to the test data. The "untuned" columns of Figure 5 shows those results. If I define "good" to mean $pd > 60 \land pf < 40\%$, then only two of my data sets (*ivy,ant*) are "good" enough for this study. Note that, as might have been expected, none of the three bottom rows of Figure 5 were "good".

Fortunately, the "tuned" columns of Figure 5 show that I can salvage some of the data sets. Pelayo and Dick [18] report that defect prediction is improved by SMOTE [19]; i.e. an oversampling of minority-class examples. Also, Fu et al. [20] report that parameter tuning with differential evolution [21] can quickly explore the tuning options of Random Forest to find better settings for the (e.g.) size of the forest, the termination criteria for tree generation, etc. The rows marked with a in Figure 5 show data sets whose performance was improved remarkably by these techniques. For example, in *poi*, the recall increased by 4% while the false alarm rate dropped by 21%.

However, as might have been expected, I could not salvage the data sets in the three bottom rows.

In summary, while I cannot trust predictors from some of my defect data sets, I can plan ways to reduce defects in *jedit, ivy, ant, lucene* and *poi*. Accordingly, when this study explores the Jureczko data, I will use these five data sets.

(Aside: One important detail to be stressed here is that, when I applied SMOTE-ing and parameter tunings, those techniques were applied to the training data and *not* the test data; i.e. I took care that no clues from the test set were ever used in this tuning process.)

## IV. Four Planning Methods

In his textbook on empirical methods, Cohen [22] advises that any supposedly better algorithm should be baselined against simpler alternatives. Accordingly, this section described XTREE (which I call Method4) against three alternatives. XTREE uses the decision tree learner of Figure 6.D. It is new to this paper. The other methods use the top-down biclustering method described in Figure 6.C which recursively divides the data in two using a dimension that captures the greatest variability in the data. I proposed Method 1 and 2 in 2012 [4] while Methods 3 comes from research conducted earlier this year [5].

Note that all methods have the properties proposed in §II. That is, they are *local learners*; i.e. different test examples will be given plans that are specialized to their particulars. Also, they are *trust-increasing*; i.e. they change examples such that they move *closer* to the training data.

### A. Methods

My description of the methods adopts the following convention. All variables set via my engineering judgement with Greek letters; e.g. $\alpha, \beta, \gamma$. In this paper, I show my current settings to these variables produces useful results. Elsewhere [20], [23], I are exploring tuning methods to find better settings but I have nothing definitive yet to report on auto-tuning planners.

*1) Method1= CD= Centroid Deltas: Summary1:* Method1 computes a plan from the difference between where you are (which I will call $C_i$) and where you want to be (which I will call $C_j$).

*Assumption1:* Method1 assumes that large data sets can be adequately represented by a few dozen (or so) centroids.

*Details1:* Method1 clusters project data by reflecting on the independent variables, then reports the delta between the cluster centroids. After clustering training data using the WHERE algorithm of Figure 6.C, Method1 replaces all clusters with a centroid $C_i$ computed from the mean/mode value of each continuous/discrete feature. After that, it finds the closest centroid $C_j$ that has a better performance score. For defect data, "better" means fewer defective examples while for the config data, "better" means lower median runtimes for the examples in that cluster. Method1 then caches the delta between the independent features between $C_i$ and $C_j$. For continuous features, this delta is $C_j - C_i$. For discrete values, this delta is the value of that feature in $C_j$. Finally, for every test case, Method1 use the distance measure $d$ shown in Figure 6.B to find the nearest centroid $C_i$. It then proposes a plan for improving that test case that is the conjunction of all the deltas between $C_i$ and $C_j$.

*2) Method2=CD+FS=Method1+Feature Selection : Summary2:* Method2 works line Method1 but now the plans only mention the $\beta = 33\%$ most informative features. Hence, Method2's plans are simpler.

*Assumption2:* Method2 assumes that, when reasoning about centroids, I can just use features that best distinguish centroids; i.e. whose values appear in just a few centroids.

*Details2:* A common result is that the signal in a table of data is mostly contained in a handful of features [24], [25]. Papakroni [26] has tested for this effect in the Jureczko data sets. Papakroni found no loss of efficacy in defect prediction after sorting all features by their information content, then making predictions using (a) all features or (b) just using 33% most informative features.

Based on the above, it might be possible to simplify the plans found by Method1 by pruning back the features in those plans. Following on from Papakroni, my Method2 returns plans containing just the top $\beta = 33\%$ most informative features. Here, "informative" means that the values of a feature are good for selecting a small set of clusters (ideally, just one). This can be estimated using the Fayyad-Iranni INFOGAIN algorithm [27] of Figure 6.E.

*3) Method3= BIC= Method2 + Best-in-cluster: Summary3:* Method3 is like Method2, but it uses more knowledge about the training data.

*Assumption3:* Method3 assumes that there exists "gradients" between and within clusters which, if used, will better guide us to finding better plans.

*Details3:* Method3 summarizes clusters into *two* examples: (1) the centroid $C_x$ found in Method1 and (2) the best-incluster $B_x$ example; i.e. the example in that cluster with the best performance score. For the Siegmud data, $B_x$ is the cluster member with the fastest runtime; for the Jureczko data, $B_x$ is the cluster member with lowest raw defect count (resolving ties at random).

Method3 connects each centroid to a nearest neighbor by *gradient*. Each gradient has a (bottom,top) end labelled ($C_i,C_j$) containing the (worst,best) centroid performance scores, respectively.. For each test instance, Method3 find the nearest gradient, the runs up to the top best end $C_j$, then extracts $B_j$ (which is the best-in-cluster associated with $C_j$). The returned plan is then computed from the delta between the test case and $B_j$.

4) *Method4=XTREE= Deltas in Decision Branches: Summary4:* Method4 builds a decision tree, then generates plans from the difference between two branches: (1) the branch to where you are and (2) the branch to where you want to be.

*Assumption4:* One potential problem with Methods 1,2 and 3 is the *unsupervised* nature of the clustering algorithm (WHERE) that executes without knowledge of the target class.

Figure 7.A: Measuring Variability
For continuous and discrete values, the *variability* can be measured using standard deviation $\sigma$ or entropy $e$. Note that where $\bar{x}$ is the mean of numeric features $x_1, x_2, ... x_n$. Also for $n$ discrete values at frequency and $p_i = f_i/N$.

Figure 7.B: Measuring distance
We use Aha et al.'s standard Euclidean distance measure [28]. For $F$ independent features, the measure returns $d(X,Y) =$

. Here, $w_i$ is a weight term for each feature (usually set to 1). Within $\Delta$, if $X_i, Y_i$ are both missing values, then return 1. Otherwise, replace any missing items with values that maximizes the following. For numerics $\Delta$ normalizes $X_i, Y_i$ (to the range 0,1 for min,max) then returns $X_i - Y_i$. For discrete variables, $\Delta$ returns 0,1 if $X_i, Y_i$ are the same,different (respectively).

Figure 7.C: Top-down Clustering with WHERE√ WHERE divides data into groups of size $\alpha = N$ using Using this measure, WHERE runs as follows:

1) Find two distance cases, $X,Y$ by picking any case $W$ at random, then setting $X$ to its most distant case, then setting $Y$ to the case most distant from $X$ (which requires only $O(2N)$ comparisons).
2) Project each case $Z$ to position $x$ on a lines running from $X$ to $Y$: if $a,b$ are distances $Z$ to $X,Y$ then $x = (a^2+c^2-b^2)/(2ac)$.
3) Split the data at the median√$x$ value of all cases.
4) For splits larger than $\alpha = $ $N$, recurse from step1.

In terms of related work, the above is similar in approach to Boley's PDDP algorithm [29], but PDDP requires an $O(N^2)$ calculation at each recursive level to find the PCA principle component. Our method, on the other hand, performs the same task with only $O(2N)$ distance calculations using the FASTMAP heuristic [30] shown in step1. Platt [31] notes that FASTMAP is a Nystrom approximation to the first component of PCA.

Figure 7.D: Top-down division with Decision Trees Find a split in the values of independent features that most reduces the variability of the dependent feature (measured using Figure 6.A). Divide data on that feature's split and for√ all splits bigger than $\alpha = N$, recurses on each split.

Figure 7.E: Finding the most informative rows
Discretize all numeric features using the Fayyad-Iranni discretizer [27] (divide numeric columns into bins $B_i$, each of which select for the fewest cluster ids). Let feature $F$ have bins $B_i$, each of which contains $n_i$ rows and let each bin $B_i$ have entropy $e_i$ computed from the frequency of clusters seen in that bin (computed from Figure 6.A). Cull the the features as per Papakroni [26]; i.e. just use the $\beta = 33\%$ most informative features where the value of feature is $F_i$ is $\sum e_i \cdot \frac{n_i}{N}$ ($N$ the number of rows).

Fig. 6. Some algorithms used in this paper.

*Supervised* methods, on the other hand, assume that it is useful to also reflect on the target class.

*Details4:* XTREE uses a supervised decision tree algorithm of Figure 6.D to divide the data. Next, XTREE builds plans from the branches of the decision trees using the code of Figure 7. That code asks three questions, the last of which returns the plan:

1) What *current* branch does a test case fall in?
2) What *desired* branch would the test case want to move to?
3) What are the *deltas* between current and desired?

V. EXPERIMENTS

This section describes an experimental design (and results) for evaluating the above four methods.

A. Experimental Design

1) *A Strategy for Evaluating Planners:* Here is my experimental design:

As shown in Figure 8, I divide the project data into two disjoint sets *train* and *test* (so *train* ∩ *test* = ∅). Next, from the train set, I build both a *planner* and a *predictor*.

My general framework does not commit to any particular choice of planner or predictor but, for the purposes of this paper:

Using the training data, divide the data using the decision tree√ of algorithm of Figure 6.D into groups of size $\alpha = N$. For each item in the test data, find the *current* leaf: take each test instance, run it down to a leaf in the decision tree. After that, find the *desired* leaf:

- Starting at *current*, ascend the tree $lvl \in \{0,1,2...\}$ levels;
- Identify *sibling* leaves; i.e. leaf clusters that can be reached from level $lvl$ that are not *current*
- Using the *score* defined above, find the *better* siblings; i.e. those with a *score* less than $\gamma = 0.5$ times the mean score of *current*. If none found, then repeat for $lvl+ = 1$. Also, return no plan if the new $lvl$ is above the root.

- Return the *closest* better sibling where distance is measured between the mean centroids of that sibling and *current*

Also, find the *delta*; i.e. the set difference between conditions in the decision tree branch to *desired* and *current*. To find that delta:

- For discrete attributes, return the value from *desired*.
- For numerics, return the numeric difference.
- For numerics discretized into ranges, return a random number selected from the low and high boundaries of the that range.

Finally, return the delta as the plan for improving the test instance.

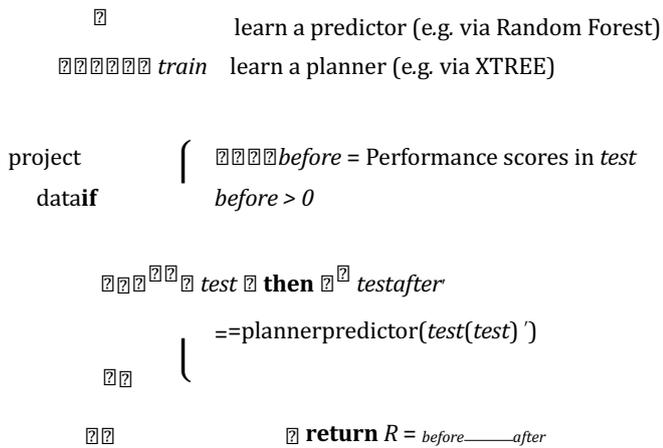

Fig. 7. XTREE

$$\text{train} \begin{cases} \text{learn a predictor (e.g. via Random Forest)} \\ \text{learn a planner (e.g. via XTREE)} \end{cases}$$

$$\text{project data if} \begin{cases} before = \text{Performance scores in } test \\ before > 0 \\ test \text{ then } test\text{after} \\ == planner\, predictor(test(test)\,') \end{cases}$$

$$\textbf{return } R = before\_\_\_\_after$$

Fig. 8. Experimental .

- My *planner* will be one of Methods 1,2,3,4;
- My *predictor* will be the Random Forest Classifier [32] (for discrete classes) and Random Forest Regressor (for continuous classes) taken from SciKit Learn [17]. I use these data miners since extensive studies have shown these to be amongst the better alternatives for mining software data [33].

As for the *test* data, this is passed to the predictor to measure performance statistics related to effectiveness.

If my predictors fail to perform effectively on the test data, then I cannot trust them to comment on my plans. Accordingly, if that performance is unsatisfactory, I abort. Recall from §III that this step indicated I should not use some of the Jureczko data.

Else, I (1) apply the planner to alter the *test* data; then (2) apply the predictor to the altered data $test'$; then (3) return data on the *before, after* predictions expressed as the ratio $R$ =

$after\ before$______. This is a unit-less ratio with the following properties:

- If $R = 1$, this means "no change from baseline";
- If $R < 1$, this indicates "some reduction to the baseline";
- If $R > 1$, this indicates "optimization failure".

*2) Statistical Methods:* My methods use some stochastic algorithms, e.g. WHERE's selection of "what example to explore first" (see Figure 6.C) and XTREE' occasional use of a random guess when deciding what part of a discretized range to include in the plan (see Figure 7). Hence, I report the $R$ values seen in 40 repeated runs (with different random number seeds). The the value 40 was chosen to be larger than the 30 samples required to satisfy the central limit theorem.

To rank my methods using the results from these 40 repeats, I use the Scott-Knott test recommended by Mittas and Angelis [34]. In that test, using the median values of each method, sort a list of $l$ = 40 values of $R$ values found in $ls$ = 4 different methods. Then, splits $l$ into sub-lists $m,n$ in order to maximize the expected value of differences in the observed performances before and after divisions. E.g. for lists $l,m,n$ of size $ls,ms,ns$ where $l = m \cup n$:

$$E(\Delta) = \frac{ms}{ls}abs(m.\mu - l.\mu)^2 + \frac{ns}{ls}abs(n.\mu - l.\mu)^2$$

Scott-Knott applies a statistical hypothesis test $H$ to check if $m,n$ are significantly different (in my case, the conjunction of A12 and bootstrapping). If so, Scott-Knott recurses on the splits. In other words, the Scott-Knott procedure being used here divides the data if *both* bootstrap sampling and effect size test agree that a division is statistically significant (with confidence of 99%) and not a small effect ($A12 \geq 0.6$).

For a justification of the use of non-parametric bootstrapping, see Efron & Tibshirani [35, p220-223]. For a justification of the use of effect size tests see Shepperd&MacDonell [36]; Kampenes [37]; and Kocaguenli et al. [38]. These researchers warn that even if an hypothesis test declares two populations to be "significantly" different, then that result is misleading if the "effect size" is very small. Hence, to assess the performance differences I first must rule out small effects using A12, a test recently endorsed by Arcuri and Briand at ICSE'11 [39].

*3) Report Format:* My results are presented by the line diagrams like those shown on the right-hand-side of the following example table. The black dot shows the median $R$ value and the horizontal likes stretch over the inter-quartile range (hereafter, IQR) that is the space from the 25th percentile value to the 75th percentile value.

| Rank | Treatment | Median | IQR | |
|------|-----------|--------|------|---|
| 1 |  |  | 0.13 | s——s |
| 2 | CD | 0.59 | 0.18 | —s— |
| 2 | BIC | 0.60 | 0.12 | —s— |
| 3 | CD+FS | 0.62 | 0.06 | = |
|  | XTREE | 0.49 |  |  |

In this example table, the rows are sorted on the median values of each method. Note that all the methods have $R < 1$ values; i.e. all these methods reduced the expected value of the performance score in that experiment while XTREE achieved the greatest reduction (down to 49% of the original value).

The above example table has a left-hand-side Rank column, computed using the Scott-Knott test described above. This column reports if the values for each method are statistically different and are more than trivially different. In this example table, CD and BIC are ranked together while XTREE and CD+FS are ranked best and worst, respectively.

*4) Other Details:* Figure 9 and Figure 10 show the effectiveness of my methods seen in 40 repeats with each data set. In these experiments, the dependent variables of Jureczko and Siegmund data sets are discrete and continuous in nature, respectively. Hence, while choosing the predictor, I used Random Forest (1) as a classifier for Jureczko data and (2) as a regressor for Siegmund data.

- For Siegmund data, I randomized the order of the data, training on one half while identifying treatment plans on the remaining test data.
- For the Jureczko data, I used the training and testing sets of Figure 5. For these datasets, all the SMOTE-ing and Random Forst tunings (discussed in §III) occurred in the *train* phase of Figure 8.

*B. Experimental Results*

Recall from my introduction that I am assessing planners on three criteria: *effectiveness*, which is how much they reduce the expected value of the changed examples; *succinctness*, which is how many things I need to change to achieve a plan; and *surprise*, which is how different the plans are from standard truisms.

*1) Effectiveness Results:* Measured in terms of effectiveness, some data sets were harder to optimize that others. SQL (in Figure 10) defied all my methods for reducing runtimes. Also, BDBJ (in Figure 10) was hard to optimize for all methods except XTREE. However, in other data sets, large reductions were observed:

- Down to 22% of the original baseline in Ant of Figure 9;
- Down to 6% of the original baseline in BDBD of Figure 10;

Overall, XTREE was most effective. It was always the topranked method and (with the exception of SQL), had significant reductions in the median performance values: median improvement of least 10% lower (i.e. better) than the next ranked method.

*2) Succinctness Results:* Figure ?? reports the percent of times in the 40 repeats that a method proposed changing a feature. The left-hand-side plot of that figure reports results from one of the Jureczko data sets (*lucene*) and the righthand-side shows a Siegmund data set (*BDBJ*).

In these plots, the *more* succinct a planning method, the *less* percent of the runs where it recommends changing a particular feature (i.e. the vertical bars in that plot are *lower*). For example, Method1 (CD) was the least succinct since it wanted to change all features (observe the change frequencies

| Rank | Treatment | Median | IQR | |
|---|---|---|---|---|
| 1 | XTREE | 0.45 | 0.1 | s |
| 2 |  | 0.55 | 0.0 | s |
| 2 | CD | 0.55 | 0.09 | s |
| 2 | CD+FS | 0.55 | 0.0 | |

Lucene

| Rank | Treatment | Median | IQR | |
|---|---|---|---|---|
| 1 | XTREE | 0.4 | 0.08 | s |

| Rank | Treatment | Median | IQR | |
|---|---|---|---|---|
| 1 |  | 0.49 | 0.13 | s |
| 2 | CD | 0.59 | 0.18 | |
| 2 | BIC | 0.6 | 0.12 | s |
| 3 | CD+FS | 0.62 | 0.06 | |

Ant

| Rank | Treatment | Median | IQR | |
|---|---|---|---|---|
| 1 |  |  |  |  |
| 2 | CD | 0.47 | 0.09 | |
| 3 | BIC CD+FS | 0.53 | 0.04 | s |
| 3 |  | 0.53 | 0.02 | s |

Ivy

| Rank | Treatment | Median | IQR | |
|---|---|---|---|---|
|  | XTREE | 0.22 | 0.19 | s s |
| 2 | CD | 0.82 | 0.07 | |
| 3 | CD+FS BIC | 0.85 | 0.0 | s |
| 3 |  | 0.88 | 0.03 | s |

Jedit

| Rank | Treatment | Median | IQR | |
|---|---|---|---|---|
| 1 | XTREE | 0.65 | 0.16 | s s |
| 2 | BIC | 0.5 | 0.04 | |
| 2 | CD+FS CD | 0.51 |  | |
| 2 |  | 0.51 | 0.04 | s s |
|  |  |  | 0.05 | |

Poi

| Rank | Treatment | Median | IQR | |
|---|---|---|---|---|
|  | BIC |  |  | s |
|  | XTREE |  |  | s s |

Fig. 9. Results on Jureczko data sets. Results from 40 repeats. Ratios of (1) number of examples with defects (expected in the test examples) after they have been altered by a planner to (2) the number of examples with defects in the original test set.

as high as 100% for all features). Method1's policy of "change everything" might be acceptable if this approach lead to the most effective changes. However, looking at Figure 9 and Figure 10, there is no evidence for this.

| Apache | | | | | LLVM | | | |
|---|---|---|---|---|---|---|---|---|
| Rank | Treatment | Median | IQR | | Rank | Treatment | Median | IQR |
| 1 | XTREE | 0.69 | 0.07 | s | 1 | XTREE | 0.88 | 0.0 s |
| 2 | BIC | 0.96 | 0.04 | | 2 | BIC | 0.97 | 0.01 |

| BDBJ | | | | | | BDBC | | | |
|---|---|---|---|---|---|---|---|---|---|
| Rank | Treatment | Median | IQR | | | Rank | Treatment | Median | IQR |
| 1 | XTREE | 0.57 | 0.08 | s | | 1 | XTREE | 0.06 | 0.06 s |
| 2 | CD+FS | 1.00 | 0.01 | | ss s | 2 | BIC | 0.34 | 0.09 s |
| 2 | CD | 1.02 | | | | 3 | CD+FS | 0.97 | 0.04 s |
| 2 | BIC | 1.04 | 0.02 | | | 4 | CD | 1.00 | 0.09 s |
| | | | 0.07 | | | | | | |

| SQL | | | | | | X264 | | | |
|---|---|---|---|---|---|---|---|---|---|
| Rank | Treatment | Median | IQR | | | Rank | Treatment | Median | IQR |
| 3 | CD | 0.98 | 0.02 | sss | | 3 | CD | 1.00 | 0.0 s ss |
| 4 | CD+FS | 1.00 | 0.0 | | | 3 | CD+FS | 1.00 | 0.0 |
| 1 | BIC | 1.00 | 0.0 | | | 2 | BIC | 0.95 | 0.02 |
| 1 | CD | 1.00 | 0.0 | | | 3 | CD | 1.00 | 0.0 |
| 1 | CD+FS | 1.01 | 0.0 | | | 3 | CD+FS | 1.01 | 0.01 |
| 1 | XTREE | 0.99 | 0.03 | s | s s s | 1 | XTREE | 0.54 | 0.03 s s ss |

Fig. 10. Results: Seigmund data sets. Results from 40 repeats. Ratios of (1) sum of software runtimes (expected in the test examples) after they have been altered by a planner to (2) the sum of the software runtimes in the original test set.

An interesting feature of Figure ?? was that fewer things were changed in the config data sets *BDBJ* than in the defect data set *lucene*. It turns out that this holds true across nearly all my data sets. Figure 11 summarizes all the change frequencies for all data sets. As with Figure ??, Fewer features were changed in the config data than in the defect prediction data. One explanation for that is the nature of the features: the defect data sets have continuous features while the config data has binary independent features (where the setting was turned "on" or "off"). When exploring these different data types, it is possible to find more
- "Gentle slopes" lead to small changes in continuous space;
- "Sharp cliffs" that lead to major change in the discrete space.

Hence, my planners can make fewer larger changes in the config data. For example, Method 3 (BIC) and Method 1 (CD) make far fewer changes in data sets with discrete features (the Siegmund data) than in data with continuous features (the Jureczko data).

As to the method that lead to best effectiveness in Figure 9 and Figure 10, I note from Figure 11 that XTREE has a consistent behavior in both the discrete and continuous data sets. Specifically, in all my data sets, XTREE changes usually changes around a fifth of the features.

*3) Surprising Results:* If a planner only ever reports conclusions that were already known, then that planner offers little value-added over "just use established wisdom".

Accordingly, I studied my results for plans that were somewhat counterintuitive.

Such a surprising plan can be seen in the *lucene* results. Recall the standard advice for OO systems: build classes that are internally cohesive with low coupling to other parts of the system [15]. I can assess the relevance of this advice to specific projects by checking how often a planner changes the coupling-related features:

- *ca*: afferent couplings = how many other classes use the specific class;
- *ce*: efferent couplings = how many other classes is used by the specific class.
- *cbm*: coupling between methods = total number of new/redefined methods to which all the inherited methods are coupled
- *cbo*: coupling between objects = a value that increases when the methods of one class access services of another.
- *ic*: inheritance coupling = number of parent classes to which a given class is coupled (includes counts of methods and variables inherited)

For the *lucene* XTREE results of Figure ??, the most frequent change was to alter the lines of code in a class (see the tallest red historgram in that figure on the *loc*, or lines of code).

Looking at the logs of my planner, I can see that XTREE's proposed change is to *reduce* the size of a class. The only way to do that while keeping the same functionality is to create a network of smaller classes that interact to produce that functionality. That is, I would need to *increase* the coupling of those classes to achieve XTREE's plan.

In theory, increasing coupling between classes complicates and confuses a class design. But the *lucene* XTREE results of Figure ?? rarely propose changing the coupling features *ca, ce, cbm, ic* (in fact, XTREE never proposes any change to *ic*).

The only coupling issue that XTREE usually adds to its plans is *cbo* (which appears 55% of the time in Figure ??). But note that this is *object* coupling measure, not class coupling. So here XTREE is warning against, say, some factory class generating a large community of agents, all of the same class, who co-ordinate on some task. This is a different issue to the class redesign issue that would be triggered by altering *loc*.

In summary, XTREE satisfies the criteria that, sometimes, it produces surprising plans. At least for the *lucense* data set, I can see advice that recommends *increasing coupling* to reduce

|  |  | XTREE | BIC | CD | CD+FS |
|---|---|---|---|---|---|
| Jureczko data | Ant | 20% | 87% | 80% | 10% |
|  | Ivy | 21% | 81% | 70% | 20% |
|  | Jedit | 19% | 92% | 90% | 20% |
|  | Lucene | 19% | 86% | 95% | 25% |
|  | Poi | 13% | 79% | 85% | 20% |
|  | mean: | 18% | 85% | 84% | 19% |
| Siegmund data | Apache | 27% | 5% | 33% | 22% |
|  | BDBJ | 7% | 8% | 3% | 15% |
|  | LLVM | 33% | 4% | 36% | 9% |
|  | X264 | 24% | 12% | 18% | 12% |
|  | BDBC | 25% | 9% | 16% | 5% |
|  | SQL | 10% | 6% | 10% | 23% |
|  | mean: | 21% | 7% | 19% | 14% |

Fig. 11. Average number of features whose values are changed by a planner.
defects.

## VI. THREATS TO VALIDITY

As with any empirical study, biases can affect the final results. Therefore, any conclusions made from this work must be considered with the following issues in mind.

### A. Learner Bias

For building the defect predictors in this study, I elected to use Random Forest and Random Forest Regressors. I chose this approach based on its reputation for having the better performance of 21 other learners for defect prediction [33]. Data mining is a large and active field, and any single study can only use a small the subset of the known classification algorithms.

That said, I have taken care to document in this paper the decisions made by engineering judgment that could affect my conclusions. The above code used a set of variables which future work should vary in order to test the internal validity of my conclusions:

- All my planners divide data into groups of size $\alpha = \sqrt{N}$
- Method2 used the top $\beta = 33\%$ most informative features (ranked using INFOGAIN);
- Method4 (XTREE) assumed that another sibling was useful if had a score less than $\gamma = 0.5$ times the mean score of the current leaf.

### B. Sampling bias

Sampling bias threatens any data mining experiment, i.e., what matters there may not be true here. For example, the data sets used here come from two sources (Seigmund et al. and Jureczko et al.) and any biases in their selection procedures threaten the validity of these results. That said, the best I can do is define my methods and publicize my data and code so that other researchers can try to repeat my results and, perhaps, point out a previously unknown bias in my analysis. Hopefully,

| Rank | Treatment | 50th | 25-75th | s |
|---|---|---|---|---|
| 1 | XTREE | 0.42 | 0.17 | s |
| 2 | CD+FS | 1.03 | 0.69 | s |
| 2 | BIC | 1.04 | 0.01 | s |
| 2 | CD | 1.11 | 1.24 |  |

Fig. 12. Results: Methods 1,2,3,4 applied to some ground-truth data (in this case, the COCOMO model). Values collected from 40 repeated runs of each method with different random seeds. Results show ratios of (1) sum of development effort (in the changes examples) to (2) the of the development effort in the original test data.

other researchers will emulate my methods in order to repeat, refute, or improve my results.

### C. Evaluation Bias

Another threat to the validity of this work is my use of predictors learned on the training data to assess the impact of my planners. This issue was discussed in detail in §II-C.

To those notes, I add a few more details. If possible, planners should be assessed via some external oracle that can accurately assess new examples. For example, in search-based software engineering, examples can be assigned objective scores via some model. In this approach, a changed example can be assessed by generating actual objective scores from the model.

For example, the COCOMO model from the University of Southern California estimates software development time using a combination of industrial data and some domain expertise from its author (Barry Boehm). Figure 12 shows results from using COCOMO as an oracle to assess my planning methods. In this experiment, random projects were used as COCOMO inputs. This generated an development time estimate for each example, which I passed to my fmy methods. 40 times, I let those methods propose changes to those projects. For assessment purposes, the changes projects were then feed back to the COCOMO oracle.

Using this approach, it is possible to assess the value of a plan by measuring its effectiveness with respect to some ground truth (in this case, the COCOMO model). As shown in Figure 12, XTREE passes this assessment. Those results from the COCOMO model endorsed the conclusions of the rest of this paper; i.e. compared to three other methods, XTREE's supervised methods are best for generating plans on how to change example projects.

## VII. Related work

### A. Planning in AI

The XTREE planner is somewhat different to the logicalbased planners explored by classical AI. That kind of planning is a logical procedure [40] that seeks an ordering on *operators* to take some domain *state* from a *start* state to a *goal* state. This classical logical approach is known to suffer from computational bottlenecks [41] while tools like XTREE will scale to any domains that can generate decision trees.

### B. Evaluating Changes

Some organizations have the resources to run repeated trials to assess different treatments. For example, in one remarkable recent study, Bente et al. report results were the same for a specification that was developed by four different organizations [42]. Given that kind of resources, it would be possible to (say) take a code base, assign it to different teams, make these teams adopt different policies, then check in 12 months' time which teams have fewer defects than the others.

Very few industrial or research groups have access to the kinds of resources needed for this kind of study (evidence: in the six years since the publication of that work, I know of only one similar study to Bente et al.). Also, given the diversity of modern software projects, it might be unreasonable to demand that all proposed changes for all projects are always evaluated by something like the Bente et al. study. Hence, this paper uses data miners to build an oracle that can assess changed examples. The advantage of this approach is that it requires far fewer resources to assess the effectiveness of proposed changes to a project.

### C. Search-based SE

Another way to propose changes to software artifacts is via some search-based method [43], [44]. Such SBSE methods are evolutionary programs that make extensive changes to some initial sample of project data (perhaps 100s to 100,000s of mutations). Each of these mutations is reassessed using some domain model. Examples of these algorithms include GALE, NSGA-II, NSGA-III, SPEA2, IBEA, particle swarm optimization, MOEA/D, etc. [23], [45]–[50].

One problem with these SBSE methods is that they can make extensive mutations to the data they are exploring. In the language of §II-C, these methods may not be *trust-increasing* those algorithms make no attempt to mutate new examples away from the kinds of data used to commission the model (in which case, I would start doubting the model's output).

Another issue with standard search-based SE methods is that they require ready access to trustworthy domain models that can offer an assessment of newly generated examples. While some domains have such models (e.g. see the COCOMO effort estimation model used in the last section), my experience is that many others do not. For example, consider software defect prediction and all the intricate issues that may lead to defects in a product. A model that includes *all* those potential issues would be very large and complex. Further, the empirical data required to validate any/all parts of that model can be hard to find.

What I would recommend is a two-pronged policy. In domains with ready access to trusted models, I recommend the kinds of tools that are widely used in search-based software engineering community such as GALE, NSGAII, NSGA-III, SPEA2, IBEA, particle swarm optimization, MOEA/D, etc. [23], [45]–[50]. Otherwise, I recommend tools like XTREE.

## VIII. Summary

The planner proposed in this paper proposes changes to software project details in order to improve the expected value of the performance scores of that part of the project. To evaluate these planners, data miners can be used to build oracles to assess planners. Such planners should be *trustincreasing*; i.e., they propose changes that generate changed examples that are closer to the training data of the data miner. One caveat here is that the evaluations I can make on the planner are only as good as the predictive performance of the data miner. Hence, if domain data does not support satisfactory predictors, then planning in that domain cannot be evaluated.

Four planners were assessed here for the tasks of reducing defects and runtimes. Three of the methods come from my prior publications [4], [5], and the conclusion of this paper is that a novel, fourth method clearly out-performs the other three (measured in terms of *effectiveness, succinctness, and surprise*). The reason for the superior performance of my new method is clear: when planning how to change to different class methods, it is best to use a learner that reflects over those classes as it divides the data.

Of course, there are many more methods for generating plans and no one paper can survey them all. However, the goal of this paper is not to claim that (e.g.) XTREE is some absolute optimal algorithm. Rather, it is to offer a baseline result (with XTREE) and an evaluation strategy that can assess if alternate methods are better than XTREE. My hope is that other researchers apply this strategy to repeat and/or improve and/or refute my results.